\documentstyle[11pt]{article}
 
\begin{document}

\baselineskip=14pt plus 0.2pt minus 0.2pt
\lineskip=14pt plus 0.2pt minus 0.2pt

\newcommand{\be}{\begin{equation}}
\newcommand{\ee}{\end{equation}}
\newcommand{\da}{\dagger}
\newcommand{\dg}[1]{\mbox{${#1}^{\dagger}$}}
\newcommand{\hlf}{\mbox{$1\over2$}}
\newcommand{\lfrac}[2]{\mbox{${#1}\over{#2}$}}

\begin{flushright}
quant-ph/9811076 \\
LA-UR-98-726 \\
\end{flushright} 

\begin{center}
\Large{\bf Time-dependent Schr\"odinger equations \\
having isomorphic symmetry algebras. \\
II. Symmetry algebras, coherent and squeezed states. \\}

\vspace{0.25in}

\large
\bigskip

Michael Martin Nieto\footnote{\noindent  Email:  
mmn@lanl.gov}\\
{\it Theoretical Division (MS-B285), Los Alamos National Laboratory\\
University of California\\
Los Alamos, New Mexico 87545, U.S.A. \\}
 
\vspace{0.25in}

 D. Rodney Truax\footnote{Email:  truax@ucalgary.ca}\\
{\it Department of Chemistry\\
 University of Calgary\\
Calgary, Alberta T2N 1N4, Canada\\}
 
\normalsize

\vskip 20pt
\today

\vspace{0.3in}

{ABSTRACT}
 
\end{center}
\begin{quotation}
\baselineskip=.33in
Using the transformations from paper I, we  show that the 
Schr\"odinger equations for: (1)
systems described by quadratic Hamiltonians,  (2) systems 
with time-varying mass,  and (3) time-dependent oscillators, all
have isomorphic Lie space-time symmetry algebras. 
The generators of the symmetry algebras are obtained 
explicitly for each case and  sets of number-operator states 
are constructed.  The algebras and the states are used to 
compute displacement-operator coherent and squeezed states.  Some 
properties of the coherent and squeezed states are calculated.  
The classical motion of these states is demonstrated.

\vspace{0.25in}


\end{quotation}

\newpage

\baselineskip=.33in

\section{ Introduction}

In this paper, we continue the investigation that  began 
in paper I {\cite{nt1}} of  three classes 
of 1-dimensional Schr\"odinger equations: 
equations with time-dependent quadratic Hamiltonians ($TQ$), 
equations with time-dependent masses ($TM$), 
and equations for time-dependent oscillators ($TO$).
They are described thusly:  

The $TQ$ class of Schr\"odinger equations, 
in units of $m=\hbar=1$, is 
\begin{eqnarray} 
S_1\Phi(x,t) & = & \{-\left[1+k(t)\right]P^2-2T +h(t)D+
           g(t)P  \nonumber\\
   &   & \hspace{1cm}-2h^{(2)}(t)X^2-2h^{(1)}(t)X
           -2h^{(0)}(t)I\}\Phi(x,t)  =  0, \label{e:pre12}
\end{eqnarray}
where $D=\lfrac{1}{2}(XP+PX)$. 

The $TM$ class of equations is 
\begin{eqnarray}
{S}_2\Theta(x,t)  =  \left\{-f(t)P^2-2T -2{f}^{(2)}
(t)X^2 -2{f}^{(1)}(t)X-2{f}^{(0)}(t)I\right\}\Theta
(x,t) = 0.\label{e:pre8}
\end{eqnarray}
In these $m=1$ units, $1/f(t)$ represents a time-dependent mass.
Rather than use the most general ($TM$) Eq. 
(\ref{e:pre8}), we shall work with the more restrictive 
(see Sec. 4.1 of paper I)
\begin{eqnarray}
\hat{S}_2\hat{\Theta}(x,t)  =  \left\{-e^{-2\nu}P^2-2T 
-2{f}^{(2)}(t)X^2 -2{f}^{(1)}(t)X-2{f}^{(0)}(t)I\right\}
\hat{\Theta}(x,t) 
  =  0,    \label{e:pre8a}  
\end{eqnarray}
\begin{equation}
f(t)=\exp[-2\nu(t)],   \label{e:pre9}
\end{equation}
where the function $\nu(t)$ will be defined below.

Finally, the time-dependent oscillator 
Schr\"odinger equations ($TO$) have the form
\begin{equation}
{S}_3\Psi(x,t') = \{-P^2+2T'-2g^{(2)}(t')X^2-2g^{(1)}(t')X-
2g^{(0)}(t')I\}\Psi(x,t')=0.\label{e:pre1}
\end{equation}

In this paper, we have four main objectives: 

First, in Section 2, we compute the 
relationship between the symmetry algebras of the three classes 
of Schr\"odinger equations, $TQ$, $TM$, and $TO$.
To do this, we start with  the 
Lie algebra of space-time symmetries associated with $TO$ equations.  
This Lie symmetry algebra for $TO$ equations is known 
{\cite{drt1}}-{\cite{nt2}} to be the Schr\"odinger algebra
\begin{equation}
{\cal SA}_1^c=su(1,1)\diamond w_1^c.\label{e:int1}
\end{equation}
Then, using the transformation developed in paper I {\cite{nt1}},
we show that 
all three classes of Schr\"odinger equations have symmetry 
algebras isomorphic to ${\cal SA}_1^c$.   

In Section 3,  
we start with the symmetry generators associated 
with $TO$ equations. We then ``work backwards" to
construct the generators of space-time transformations for first
$TM$ and then $TQ$.  These calculations constitute the second objective. 
  
In Section 4, working with the oscillator subalgebras of 
$({\cal SA})_1^c$ for each class of Schr\"odinger equations, 
we obtain a set of solutions for each.  This has already 
been done for $TO$ {\cite{drt2,nt2,gt1}}.  Here we extend 
the method to the $TM$ and $TQ$ equations.  This  
completes the third objective. 

The fourth objective   
concerns the calculation of coherent states (Section 5) 
and squeezed states (Section 6) for the three classes of equations.  
It is natural, in this 
context, to take advantage of the Lie symmetry to construct 
displacement-operator coherent states (DOCS) 
\cite{gt1}-\cite{amp1} and displacement-operator squeezed states 
(DOSS) {\cite{nt2,fns1,nt3}}. 
We make extensive use of the isomorphism of the symmetry 
algebras and the results of Section 4 to calculate properties 
of the coherent (Section 5) and squeezed (Section 6) states.  

We close with comments on uncertainty relations and the classical 
equations of motion. 

Elsewhere \cite{else}, we shall apply the general 
results of paper I and this article to the calculation of space-time 
number-state, coherent-state, and squeezed-state wave functions for 
$TM$ systems that have been studied by others. 
We shall give detailed accounts of the solutions. 


\section{Lie Symmetries}

Starting with the $TQ$ Schr\"odinger equation (\ref{e:pre12})
and using the definitions in Eq. (I-6), we express its 
Lie symmetries as {\cite{drt1,drt2,nt2,gt1,wm1}}  
\begin{equation}
L_1 = -iA_1(x,t)T+iB_1(x,t)P+C_1(x,t)I.
\label{e:sym12a}
\end{equation}
For $L_1$ to be a Lie symmetry of Eq. (\ref{e:pre12}), it must satisfy 
the operator equation
\begin{equation}
[S_1,L_1]=\lambda_1(x,t)S_1,\label{e:sym12b}
\end{equation}
where $S_1$ is the $TQ$ Schr\"odinger operator from Eq. 
(\ref{e:pre12}).  The function $\lambda_1$ depends on the 
variables $x$ and $t$.  
As a consequence, if $\Phi(x,t)$ is a solution of 
the $TQ$ equation (\ref{e:pre12}), then $L_1\Phi(x,t)$ is also 
a solution to this equation. 

Next, denote the Lie symmetries of the $TM$ Schr\"odinger 
equation (\ref{e:pre8a}) by 
\begin{equation}
\hat{L}_2 = -i\hat{A}_2(x,t)T+i\hat{B}_2(x,t)P
+\hat{C}_2(x,t)I.\label{e:sym8a}
\end{equation}
If $\hat{S}_2$ is the $TM$ Schr\"odinger operator given in 
Eq. (\ref{e:pre8a}), then for $\hat{L}_2$ to be a symmetry of Eq.
(\ref{e:pre8a}), it must satisfy the commutator relation
[$\hat{\lambda}_2$ is a function of $x$ and $t$]
\begin{equation}
[\hat{S}_2,\hat{L}_2]=\hat{\lambda}_2(x,t)\hat{S}_2.\label{e:sym8b}
\end{equation}

Finally, for the $TO$ Schr\"odinger equation (\ref{e:pre1}), 
its Lie symmetries {\cite{drt1,drt2,nt2,gt1,wm1}} are 
\begin{equation}
{L}_3 = -i{A_3}(x,t')T'+i{B_3}(x,t')P+{C_3}(x,t')I,
\label{e:sym1a}
\end{equation}
where $T'=i\partial_{t'}$.
For the operator, ${L}_3$, to be a symmetry of Eq. (\ref{e:pre1}),
it must satisfy 
\begin{equation}
[{S}_3,{L}_3]=\lambda_3(x,t'){S}_3,\label{e:sym1b}
\end{equation} 
where ${S}_3$, is the Schr\"odinger operator given in Eq. 
(\ref{e:pre1}) and $\lambda_3$ is a function of $x$ and $t'$ (not $t$).  

To obtain the coefficients of the operators $T$ or $T'$, $P$, 
and $I$, in Eqs. (\ref{e:sym12a}), (\ref{e:sym8a}), and 
(\ref{e:sym1a}), we could substitute these operators into 
Eqs. (\ref{e:sym12b}), (\ref{e:sym8b}), and (\ref{e:sym1b}), 
respectively, and solve the three sets of coupled partial 
differential equations for the coefficients of $T$ or $T'$, $P$, 
and $I$.  This has been done elsewhere for the $TO$ class of 
Schr\"odinger equations {\cite{drt1}}-{\cite{gt1}}.  

However, here
we shall adopt a different approach.  
Our present  objective  is to establish a connection 
between the Lie symmetries of the  
$TQ$, $TM$, and $TO$ equations; (\ref{e:pre12}), 
(\ref{e:pre8a}), and (\ref{e:pre1}), respectively. 
We achieve this by starting with the 
$TO$ symmetries (\ref{e:sym1a}) and transforming them into the 
$TM$ symmetries.  Then, we obtain the $TQ$ symmetries 
from the $TM$ symmetries.

In the first step, we transform from the $(x,t')$ 
to the $(x,t)$ coordinate system,
taking us from $TO$ to $TM$.  Making use of 
$1=e^{2\nu}(\partial t'/\partial t) \equiv e^{2\nu} f(t)$, 
Eq. (\ref{e:sym1b}) becomes
\begin{equation}
[\hat{\cal S}_2,\hat{L}]=\hat{\lambda}(x,t)\hat{\cal S}_2,
\label{e:sym16}
\end{equation}
where $\hat{\lambda}(x,t)=(\lambda_3\circ t')(x,t)$.  The 
generator, $\hat{L}$, takes the form
\begin{equation}
\hat{L}=-i\hat{A}(x,t)T+i\hat{B}(x,t)P+\hat{C}
(x,t)I,\label{e:sym20}
\end{equation}
\begin{equation}
\hat{A}(x,t)= ({A_3}\circ t')(x,t)e^{2\nu},~~~
\hat{B}(x,t)=({B_3}\circ t')(x,t),~~~\hat{C}(x,t)
=({C_3}\circ t')(x,t).\label{e:sym24}
\end{equation}

\indent From  $\hat{S}_2=e^{-2\nu}\hat{\cal S}_2$,
we obtain
\begin{equation}
[e^{2\nu}\hat{S}_2,\hat{L}]=\hat{\lambda}(x,t)e^{2\nu}
\hat{S}_2, \label{e:sym28}
\end{equation}
which, after rearranging, yields
\begin{equation}
[\hat{S}_2,\hat{L}]=\left(\hat{\lambda}+e^{-2\nu}[\tilde{L},e^{2\nu}]
\right)\hat{S}_2=\left(\hat{\lambda}(x,t)+2\hat{A}
{{d\nu}\over{dt}}\right)\hat{S}_2,\label{e:sym36}
\end{equation}
where $\hat{A}$ is given in Eq. (\ref{e:sym24}).  Therefore, 
$\hat{L}$ is a symmetry of $\hat{S}_2$.  Comparing 
Eqs. (\ref{e:sym8b}) and (\ref{e:sym36}), we can identify $\hat{L}_2$ 
with $\hat{L}$ if
\begin{equation}
\hat{\lambda}_2=\hat{\lambda}+2\hat{A}
{{d\nu}\over{dt}},\label{e:sym44}
\end{equation}
\begin{equation}
\hat{A}_2(x,t)=\hat{A}(x,t),~~~~~\hat{B}_2(x,t)=
\hat{B}(x,t),~~~~~\hat{C}_2(x,t)=\hat{C}(x,t).
\label{e:sym48}
\end{equation}
This means that $\hat{L}$ is a symmetry of both Eq. (\ref{e:pre8a}) 
and $\hat{\cal S}_2\hat{\Theta}(x,t)=0$, but with different ``lambda'' 
functions; $\hat{\lambda}_2$ and $\hat{\lambda}$, respectively.

In the final step, we transform the commutator bracket (\ref{e:sym8b}) 
with the transformation $R(\mu,\nu,\kappa)$ of Eq. (I-14),
thereby going from $TM$ to $TQ$.  
Inverting the transformation, 
$\hat{S}_2=R(\mu,\nu,\kappa)S_1R^{-1}(\mu,\nu,\kappa)$, 
we obtain the commutator (\ref{e:sym12b}), where 
\begin{equation}
S_1 = R^{-1}(\mu,\nu,\kappa)\hat{S}_2R(\mu,\nu,\kappa),~~~L_1= 
R^{-1}(\mu,\nu,\kappa)\hat{L}_2R(\mu,\nu,\kappa),\label{e:sym52}
\end{equation}
\begin{eqnarray}
\lambda_1(x,t) & = & R^{-1}(\mu,\nu,\kappa)\hat{\lambda}_2
R(\mu,\nu,\kappa)  \nonumber\\*[1mm]
 & = & R^{-1}(\mu,\nu,\kappa)\left(\hat{\lambda}+2\hat{A}
{{d\nu}\over{dt}}\right)R(\mu,\nu,\kappa). \label{e:sym56}
\end{eqnarray} 
Here, $d\nu/dt$ is given by Eq. (I-32). 

For $TO$ Schr\"odinger equations (\ref{e:pre1}), both 
${A}_3$ and $\lambda_3$ are functions of $t'$ only 
\cite{drt1}-\cite{kt1}.  Therefore, $\hat{\lambda}$ is a function 
of $t$ only and, according to Eq. (\ref{e:sym44}), 
$\hat{\lambda}_2$ is a function of $t$ only. 
Since the transformation $R(\mu,\nu,\kappa)$ involves no time 
derivatives,  we have 
\begin{equation}
\lambda_1(t)=\hat{\lambda}_2(t)=\hat{\lambda}(t)+
\left(8h^{(2)}(t)\kappa-h(t)\right)\hat{A}.
\label{e:sym60}
\end{equation}


\section{Lie Symmetries for Each Class of Schr\"odinger equation}


\subsection{TO Symmetries}

The six generators of Lie space-time symmetries for the 
$TO$ Schr\"odinger equation (\ref{e:pre1}) have been calculated 
previously \cite{drt1}-\cite{kt1}.  They form a basis for the 
Lie algebra $sl(2,{\bf R})\Box w_1$ {\cite{drt1,kt1}}.  We prefer 
to use its complexification {\cite{drt2,kt1}}, which we have called 
the Schr\"odinger algebra, denoted by ${\cal SA}_1^c$ in Eq. 
(\ref{e:int1}).  We shall work with the generators 
of ${\cal SA}_1^c$ only.  

First, the three generators which form a 
basis for the Heisenberg-Weyl subalgebra, $w_1^c$, are  
\begin{equation}
  {J}_{3-}=i\left\{\xi P-X\dot{\xi}+{\cal C}I\right\},~~~~~
{J}_{3+}=i\left\{-\bar{\xi} P+X\dot{\bar{\xi}}-{\bar{\cal C}}I
\right\}, ~~~~~ I = 1.  \label{e:to4}
\end{equation}
Both $\xi$ and its complex conjugate, $\bar{\xi}$, are functions of 
$t'$ and are two linearly independent solutions of the 
second-order, ordinary differential equation \cite{drt1}-\cite{kt1}
\begin{equation}
\ddot{\gamma}+2g^{(2)}(t')\gamma=0.\label{e:to8}
\end{equation}
The Wronskian of the two solutions is a constant,  
\begin{equation}
W(\xi,\bar{\xi})=\xi\dot{\bar{\xi}}-\dot{\xi}\bar{\xi}=-i.
\label{e:to9}
\end{equation}
A dot over the function indicates differentiation by $t'$, 
i.e. $\dot{\xi}=d\xi/dt'$.
The function ${\cal C}t')$ is
\begin{equation}
{\cal C}(t') = c(t')+{\cal C}^o, ~~~~~~~~
c(t')=\int_{t'_o}^{t'}ds\,g^{(1)}(s)\xi(s),
\label{e:to12}
\end{equation}
where ${\cal C}^o$ is an integration constant {\cite{kt1}}.

The three generators of the $su(1,1)$ subalgebra have the form
\begin{eqnarray}
 & {M}_{3-} = -\left\{-\phi_1T'+\lfrac{1}{2}\dot{\phi}_1D+{\cal E}_1P
-\lfrac{1}{4}\ddot{\phi}_1X^2-\dot{\cal E}_1X+({\cal D}_1+g^{(0)}
\phi_1)I\right\}, & \nonumber\\
 & {M}_{3+} = -\left\{-\phi_2T'+\lfrac{1}{2}\dot{\phi}_2D+{\cal E}_2P
-\lfrac{1}{4}\ddot{\phi}_2X^2-\dot{\cal E}_2X+({\cal D}_2+g^{(0)}
\phi_2)I\right\}, & \nonumber\\
 & {M}_3 = -\left\{-\phi_3T'+\lfrac{1}{2}\dot{\phi}_3D+{\cal E}_3P
-\lfrac{1}{4}\ddot{\phi}_3X^2-\dot{\cal E}_3X+({\cal D}_3+g^{(0)}
\phi_3)I\right\}. & \label{e:to16}
\end{eqnarray}
The three functions, $\phi_j$, $j=1,2,3$, 
are defined as  (note that $\phi_3$ is a real function of $t'$)
{\cite{drt2,kt1}} 
\begin{equation}
\phi_1=\xi^2,~~~~\phi_2=\bar{\xi}^2,~~~~\phi_3=2\xi\bar{\xi}.
\label{e:to20}
\end{equation}

The remaining $t'$-dependent functions are defined in terms of $\xi$, 
$\bar{\xi}$,   ${\cal C}$, and  $\bar{\cal C}$
\begin{eqnarray}
 & {\cal E}_1=-\xi{\cal C},~~~{\cal E}_2=-\bar{\xi}\bar{\cal C},~~~
{\cal E}_3=-\xi\bar{\cal C}-\bar{\xi}{\cal C}, & 
\label{e:to24a}\\*[1mm]
 & {\cal D}_1=-\lfrac{1}{2}{\cal C}^2,~~~{\cal D}_2=-\lfrac{1}{2}
\bar{\cal C}^2,~~~{\cal D}_3=-{\cal C}\bar{\cal C}. & 
\label{e:to24b}
\end{eqnarray}
Both ${\cal E}_3$ and ${\cal D}_3$ are real functions of $t'$.


\subsection{$TM$ Symmetries from $TO$ Symmetries}

First, we calculate the generators, $\hat{J}_{2\pm}$.  In the initial 
step, we transform the operators ${J}_{3\pm}$ from the $(x,t')$ 
coordinate system to the $(x,t)$ system, as described in 
Sections 3 and 4.1 of paper I:
\begin{equation}
\hat{J}_{2-}  =  i\left\{\hat{\xi}P-X\hat{\dot{\xi}}+ \hat{\cal C}
I\right\},~~~~~~~
\hat{J}_{2+}  =  i\left\{-\hat{\bar{\xi}}P+X\hat{\dot{\bar{\xi}}}
- \hat{\bar{\cal C}}I\right\},  \label{e:tme1}
\end{equation}
\begin{eqnarray}
  \hat{\xi}(t)=(\xi\circ t')(t),~~~~~~~\hat{\dot{\xi}}(t)
=(\dot{\xi}\circ t')(t), 
~~~~~~~~
  \hat{\cal C}(t)=({\cal C}\circ t')(t)=(c\circ t')(t)
+{\cal C}^o.   \label{e:tme4b}
\end{eqnarray}
In Eq. (\ref{e:tme4b}) we have used definition (\ref{e:to12}) for ${\cal C}$.

It is important to keep in mind that, in general,
\begin{equation}
{{d}\over{dt}}\hat{\xi}(t)\ne \hat{\dot{\xi}}(t).
\label{e:tme8}
\end{equation}
The `dot' over a function will always indicate differentiation by 
$t'$.  Differentiation by $t$ will always be written 
in full notation.  Also, an important relationship between $\hat{\xi}$ 
and $\hat{\bar{\xi}}$ is 
\begin{equation}
\hat{\xi}\hat{\dot{\bar{\xi}}}-\hat{\dot{\xi}}\hat{\bar{\xi}}=-i,
\label{e:tme10}
\end{equation}
which follows from the Wronskian (\ref{e:to9}) and the definitions 
in Eq (\ref{e:tme4b}).

Now, we proceed in the same way as before 
to obtain the operators spanning the 
$su(1,1)$ algebra.  First, because of Eq. (I-51) and $f(t)=e^{-2\nu}$,
we note that 
\begin{equation}
T'=e^{2\nu}T. \label{e:tme12}
\end{equation}
The three generators of $su(1,1)$ have the form 
\begin{eqnarray}
 & \hat{M}_{2-} = -\left\{-\hat{\phi}_1e^{2\nu}T+\lfrac{1}{2}
\hat{\dot{\phi}}_1D+\hat{E}_1P
-\lfrac{1}{4}\hat{\ddot{\phi}}_1X^2-\hat{\dot{E}}_1X
+(\hat{D}_1+\hat{g}^{(0)}\hat{\phi}_1)I\right\}, & \nonumber\\
 & \hat{M}_{2+} = -\left\{-\hat{\phi}_2e^{2\nu}T+\lfrac{1}{2}
\hat{\dot{\phi}}_2D+\hat{E}_2P
-\lfrac{1}{4}\hat{\ddot{\phi}}_2X^2-\hat{\dot{E}}_2X
+(\hat{D}_2+\hat{g}^{(0)}\hat{\phi}_2)I\right\}, & \nonumber\\
 & \hat{M}_2 = -\left\{-\hat{\phi}_3e^{2\nu}T+\lfrac{1}{2}
\hat{\dot{\phi}}_3D+\hat{E}_3P
-\lfrac{1}{4}\hat{\ddot{\phi}}_3X^2-\hat{\dot{E}}_3X
+(\hat{D}_3+\hat{g}^{(0)}\hat{\phi}_3)I\right\}, & 
\label{e:tme16}
\end{eqnarray}
where, for $j=1,2,3$ and again keeping in mind the comment on `dots'
following Eq. (\ref{e:tme8}),
\begin{eqnarray}
 & \hat{D}_j(t)=({\cal D}_j\circ t')(t),  ~~~~~~~~
\hat{E}_j(t)=({\cal E}_j\circ t')(t), ~~~~~~~~
\hat{\dot{E}}_j(t)=(\dot{\cal E}_j\circ t')(t),& 
\nonumber\\*[1mm]
 & \hat{\phi}_j(t)=(\phi_j\circ t')(t),~~~~~~~~
\hat{\dot{\phi}}_j(t)=(\dot{\phi}_j\circ t')(t),~~~~~~~~
\hat{\ddot{\phi}}_j(t)=(\ddot{\phi}_j\circ t')(t). & 
\label{e:tme20}
\end{eqnarray}
Also, according to Eq. (I-70), we have 
\begin{eqnarray}
\hat{g}^{(0)}(t)  =  (g^{(0)}\circ t')(t)  
  =  h^{(0)}(t)e^{2\nu}+h^{(1)}(t)e^{3\nu}\mu+
h^{(2)}(t)e^{4\nu}\mu^2 
  =  e^{2\nu}f^{(0)}(t).    \label{e:tme24}
\end{eqnarray}


\subsection{$TQ$ Symmetries from $TM$ Symmetries}

A basis for the Lie symmetries associated with the $TQ$ Schr\"odinger 
equation can be obtained by applying the transformation 
in Eq. (\ref{e:sym52}) to the generators for the $TM$ class of Lie 
symmetries obtained in the previous subsection.  From Eq. (\ref{e:tme1}), 
and using (see Ref. {\cite{wm2}}) 
\begin{equation}
e^{A}Be^{-A}=B+[A,B]+\lfrac{1}{2!}[A,[A,B]]+\cdots,\label{e:tqh1}
\end{equation}
we obtain 
\begin{eqnarray}
 & J_{1-} = R^{-1}(\mu,\nu,\kappa)\hat{J}_{2-}R(\mu,\nu,\kappa) 
= i\left\{\Xi_PP-X\Xi_X+\Xi_II\right\}, & \label{e:tqh4a}\\*[2mm]
 & J_{1+} = R^{-1}(\mu,\nu,\kappa)\hat{J}_{2+}R(\mu,\nu,\kappa)
= i\left\{-\bar{\Xi}_PP+X\bar{\Xi}_X-\bar{\Xi}_II\right\},  &
\label{e:tqh4b}
\end{eqnarray}
\begin{eqnarray}
 & \Xi_P(t) = \hat{\xi}(t)e^{\nu}+2\hat{\dot{\xi}}(t)
\kappa e^{-\nu},~~~~~~~\Xi_X(t)=
\hat{\dot{\xi}}(t)e^{-\nu}, ~~~~~~~
\Xi_I(t) = \hat{\cal C}(t)+\mu\hat{\dot{\xi}}(t).
 & \label{e:tqh8a}
\end{eqnarray}
The analogue of Eqs. (\ref{e:to9}) and (\ref{e:tme10}) is
\begin{equation}
\Xi_P\bar{\Xi}_X-\bar{\Xi}_P\Xi_X=-i.\label{e:tqh10}
\end{equation}

To obtain the basis of the $su(1,1)$ Lie subalgebra, we observe that 
\begin{eqnarray}
 & M_{1-}=R^{-1}(\mu,\nu,\kappa)\hat{M}_{2-}R(\mu,\nu,\kappa),~~~
M_{1+}=R^{-1}(\mu,\nu,\kappa)\hat{M}_{2+}R(\mu,\nu,\kappa), & 
\nonumber\\*[1mm]
 & M_{1}=R^{-1}(\mu,\nu,\kappa)\hat{M}_{2}R(\mu,\nu,\kappa). &
\label{e:tqh12}
\end{eqnarray}
Keeping in mind that the functions, $\mu$, $\nu$, 
and $\kappa$ are $t$-dependent, we find that  
\begin{eqnarray}
M_{1-} & = & -\left\{-C_{1,T}T+C_{1,P^2}P^2+C_{1,D}D+C_{1,P}P
+C_{1,X^2}X^2+C_{1,X}X +C_{1,I}I\right\},\nonumber\\*[1mm]
M_{1+} & = & -\left\{-C_{2,T}T+C_{2,P^2}P^2+C_{2,D}D+C_{2,P}P
+C_{2,X^2}X^2+C_{2,X}X +C_{2,I}I\right\},\nonumber\\*[1mm]
M_1 & = & -\left\{-C_{3,T}T+C_{3,P^2}P^2+C_{3,D}D+C_{3,P}P
+C_{3,X^2}X^2+C_{3,X}X +C_{3,I}I\right\},\label{e:tqh16}
\end{eqnarray}
where, for $j=1,2,3$, the coefficients are
\begin{eqnarray}
C_{j,T} & = & \hat{\phi}_j(t)e^{2\nu},\label{e:tqh20a}
\\*[1mm]
C_{j,P^2} & = & \hat{\phi}_j(t)\left(\lfrac{1}{2}k(t)-
4h^{(2)}(t)\kappa^2\right)e^{2\nu}-\hat{\dot{\phi}}_j(t)
\kappa-\hat{\ddot{\phi}}_j(t)\kappa^2e^{-2\nu},
\label{e:tqh20b}\\*[1mm]
C_{j,D} & = & \hat{\phi}_j(t)\left(-\lfrac{1}{2}h(t)+
4h^{(2)}(t)\kappa\right)e^{2\nu}+\lfrac{1}{2}\hat{\dot{\phi}}_j
(t)+\hat{\ddot{\phi}}_j(t)\kappa e^{-2\nu}, \label{e:tqh20c}
\end{eqnarray}
\begin{eqnarray}
C_{j,P} & = & \hat{\phi}_j(t)\left(-\lfrac{1}{2}
g(t)+2h^{(1)}(t)\kappa\right)e^{2\nu}
+\hat{E}_j(t)e^{\nu}\nonumber\\
    &   & \hspace{1cm} -\lfrac{1}{2}\hat{\dot{\phi}}_j(t)
\mu e^{\nu}-\hat{\ddot{\phi}}_j(t)\kappa\mu e^{-\nu}
+2\hat{\dot{E}}_j(t)\kappa e^{-\nu},\label{e:tqh20d}\\*[1mm]
C_{j,X^2} & = & -\lfrac{1}{4}\hat{\ddot{\phi}}_j(t)e^{-2\nu}, 
\label{e:tqh20e}\\*[1mm]
C_{j,X} & = & -\hat{\dot{E}}_j(t)e^{-\nu}+\lfrac{1}{2}
\hat{\ddot{\phi}}_j(t)\mu e^{-\nu},\label{e:tqh20f}\\*[1mm]
C_{j,I} & = & \hat{\cal D}_j(t)-\lfrac{1}{4}\hat{\ddot{\phi}}_j
(t)\mu^2+\hat{\dot{E}}_j(t)\mu\nonumber\\
 &   & \hspace{1cm} +\hat{\phi}_j(t)\left(h^{(0)}(t)
+h^{(1)}(t)\mu e^{\nu}+h^{(2)}(t)\mu^2 e^{2\nu}\right)e^{2\nu}.
\label{e:tqh20g}
\end{eqnarray}

\subsection{Commutation relations and algebraic structure}

The commutation relations for the symmetry operators have been 
worked out previously \cite{drt2}-\cite{nt2} and the structure of 
the Lie algebra is known to be $su(1,1)\Box w_1^c$.  The nonzero 
$TO$ commutators are as follows:
\vspace{2mm}
\noindent For the $w_1^c$ subalgebra:
\begin{equation}
[{J}_{3-},{J}_{3+}]=I,\label{e:to28}
\end{equation}
\vspace{1mm}
\noindent For the $su(1,1)$ subalgebra:
\begin{equation}
[M_{3+},M_{3-}]=-M_3,~~~[M_3,M_{3-}]=
-2M_{3-},~~~[M_3,M_{3+}]=+2M_{3+}.
\label{e:to32}
\end{equation}
The nonzero commutators involving operators from each of 
the two subalgebras are
\begin{eqnarray}
 & [M_3,J_{3-}]=-J_{3-},~~~~[M_3,J_{3+}]
=+J_{3+}, & \nonumber\\
 & [M_{3-},J_{3+}]=-J_{3-},~~~~[M_{3+},J_{3-}]=
+J_{3+}. & \label{e:to36}
\end{eqnarray}

Since commutation relations are preserved by each segment of the 
transformation ($TO\rightarrow TM$) and ($TM\rightarrow TQ$), 
the Lie algebras of operators associated with  
$TM$ Schr\"odinger equations and $TQ$ Schr\"odinger equations 
are isomorphic to $su(1,1)\Box w_1^c$.  We take advantage of this 
isomorphism to define a set of generic operators, 
$\{M,M_{\pm},J_{\pm},I\}$, where the subset $\{M,M_{\pm}\}$ forms a 
subalgebra with the $su(1,1)$ structure:
\begin{equation}
[M_+,M_-]=-M,~~~~~~[M,M_{\pm}]=\pm 2M_{\pm},\label{e:gnc1}
\end{equation}
and the subset $\{J_{\pm},I\}$ forms a subalgebra with the 
$w_1^c$ structure:
\begin{equation}
[J_-,J_+]=I.\label{e:gnc4}
\end{equation}
The nonzero commutation relations between operators from each of the two 
subalgebras are
\begin{equation}
[M,J_{\pm}]=\pm J_{\pm},~~~~~[M_-,J_+]=-J_-,~~~~~[M_+,J_-]= +J_+. 
\end{equation}
The operators $M$ and $M_{\pm}$ are identified with $M_j$ and $M_{j\pm}$, 
respectively, and the $J_{\pm}$ are identified with $J_{j\pm}$. 
This is for $j=1,2,3$, with or without hats.


\section{Eigenstates of the Number Operator}

\subsection{Casimir Operators}

In the following analysis, we do not require the operators 
$M_{\pm}$.  We shall consider only the subalgebra 
consisting of the operators $\{M,J_{\pm},I\}$, 
satisfying the nonzero commutation relations
\begin{equation}
[M,J_{\pm}]=\pm J_{\pm},~~~~[J_-,J_+]=I,\label{e:eig1}
\end{equation}
and its representation spaces.  Regardless of the system 
we are working with, we refer to this subalgebra as 
the oscillator subalgebra, denoted by $os(1)$, 
with one Casimir operator
\begin{equation}
{\bf C}=J_+J_--MI.\label{e:eig2}
\end{equation}
              
\indent For the $TQ$ class of equations, we have the expressions 
\begin{equation}
J_{1+}J_{1-} =-\lfrac{1}{2}\hat{\phi}_3(t)e^{2\nu}S_1+M_1-\lfrac{1}{2},
\label{e:eig20}
\end{equation} 
\begin{equation}
{\bf C}_1=J_{1+}J_{1-}-M_1I =-\lfrac{1}{2}\hat{\phi}_3(t)
e^{2\nu}S_1-\lfrac{1}{2},\label{e:eig24}
\end{equation}
where $S_1$ is the $TQ$ Schr\"odinger operator in Eq. 
(\ref{e:pre12}).  The operators, $J_{1\pm}$ are given in Eqs. 
(\ref{e:tqh4a}) and (\ref{e:tqh4b}) and $M_1$ 
is found in Eq. (\ref{e:tqh16}).

The expression analogous to Eq. 
(\ref{e:eig20}) for the $TM$ class of equations is 
\begin{equation}
\hat{J}_{2+}\hat{J}_{2-} =-\lfrac{1}{2}\hat{\phi}_2(t)e^{2\nu}\hat{S}_2
+\hat{M}_2-\lfrac{1}{2},
\label{e:eig12}
\end{equation}
where $\hat{S}_2$ is the $TM$ Schr\"odinger operator from Eq. 
(\ref{e:pre8a}).  The operators $\hat{J}_{2\pm}$ and $\hat{M}_2$, 
defined in Eqs. (\ref{e:tme1}) and (\ref{e:tme16}), are members of the 
$TM$ $os(1)$ algebra.  Its Casimir operator is 
\begin{equation}
\hat{\bf C}_2=\hat{J}_{2+}\hat{J}_{2-}-\hat{M}_2I =
-\lfrac{1}{2}\hat{\phi}_3(t)e^{2\nu}\hat{S}_2-\lfrac{1}{2}.
\label{e:eig16}
\end{equation}

Similarly, as shown in Refs. {\cite{drt2,nt2}}, 
the Casimir operator for the $TO$ $os(1)$ Lie algebra is 
\begin{equation}
{\bf C}_{3}={J}_{3+}{J}_{3-} -{M}_3I =-\lfrac{1}{2}
\phi_3(t'){S}_3-\lfrac{1}{2}.\label{e:eig4}
\end{equation}
The second equality follows from the relationship
\begin{equation}
{J}_{3+}{J}_{3-} =-\lfrac{1}{2}\phi_3(t'){S}_3
+{M}_3-\lfrac{1}{2}.\label{e:eig8}
\end{equation}
where ${S}_3$ is the $TO$ Schr\"odinger operator from 
Eq. (\ref{e:pre1}).  The operators ${J}_{3\pm}$ and ${M}_3$, 
defined in Eqs. (\ref{e:to4}) and (\ref{e:to16}), are members of the 
$TO$ $os(1)$ algebra.  
                         
\subsection{Number States}

Previously, we showed {\cite{drt2}} that certain states of the 
time-dependent oscillator equation (\ref{e:pre1}) form a 
representation space for the oscillator algebra, $os(1)$.  
Since the representation spaces depend primarily upon the 
algebraic structure of the algebra, the same will hold 
true for the $TM$ and $TQ$ equations.  
In this representation, the operators $M$ and 
${\bf C}$ are diagonal.  If ${\bf Z}_0^+$ denotes 
the set of nonnegative integers and if we denote the spectrum 
of the operator $M$ by ${\rm{Sp}}(M)$, then
\begin{equation}
{\rm{Sp}}(M)=\left\{n+\lfrac{1}{2},\, n\in {\bf Z}^+_0
\right\},\label{e:eig28}
\end{equation}
and the spectrum is bounded below.  Let 
$\{\Omega_n,\,n\in {\bf Z}_0^+\}$ be a basis for this representation 
space.  Each vector $\Omega_n$ in this set is an eigenvector of 
the operator $M$ with eigenvalue $n+1/2$.  

The extremal state, $\Omega_0$, is annihilated by the operator 
$J_-$, that is
\begin{equation}
J_-\Omega_0=0.\label{e:eig32}
\end{equation}
Furthermore, by requiring that each $\Omega_n$ be a solution to 
an appropriate Schr\"odinger equation in each class, Eq. 
(\ref{e:eig4}) implies that, for all $n\in {\bf Z}_0^+$, 
\begin{equation}
{\bf C}\Omega_n = -\lfrac{1}{2}\Omega_n. \label{e:eig36}
\end{equation}

The action of the basis of the 
$TO$ Lie algebra on the vectors in the representation space is 
\begin{eqnarray}
 & M\Omega_n=\left(n+\lfrac{1}{2}\right)\Omega_n, & 
\label{e:eig40a}\\*[1mm]
 & J_+\Omega_n=\sqrt{n+1}\Omega_{n+1},~~~~~J_-\Omega_n=\sqrt{n}
\Omega_{n-1}, & \label{e:eig40b}
\end{eqnarray}
for $n\in {\bf Z}_0^+$.  We indicate this irreducible representation 
by the symbol $\uparrow_{-1/2}$, where the subscript is the 
eigenvalue of the Casimir operator ${\bf C}$.  From the 
extremal state, $\Omega_0$, we can obtain all higher-order states by 
repeated application of the raising operator, $J_+$:
\begin{equation}
\Omega_n=\sqrt{\lfrac{1}{n!}}\left(J_+\right)^n\Omega_0.
\label{e:eig44}
\end{equation}

The states $\Omega_n$ are also eigenstates of the operator 
$J_+J_-$ with eigenvalue $n$.  We shall refer to 
the eigenfunctions $\Omega_n$ as number states.  
We emphasize that the number states are generally not 
eigenfunctions of a Hamiltonian.  Therefore, we do not refer to the 
extremal state, $\Omega_0$, as a ground state nor to the higher-order 
states as excited states.  We reserve the terms `ground state' and 
`excited state' for states that are energy 
eigenstates of the system.

\begin{center}
\begin{tabular}{|r|cccc|c|}
\multicolumn{1}{l}{Table 1.} & 
\multicolumn{5}{l}{Generic symbols and their values according to class.
}\\*[1.5mm] \hline\hline
 &  &  &  &  & \\*[-.25cm]
     & $M$ & ${\bf C}$ & $J_+$ & $J_-$ & $\Omega_n$ \\*[1.5mm] \hline
 &  &  &  &  & \\*[-.25cm]
$TQ$ & $M_1$ & ${\bf C}_1$ & $J_{1+}$ & $J_{1-}$ & $\Phi_n$ \\*[1mm]
$TM$ & $\hat{M}_2$ & $\hat{\bf C}_2$ & $\hat{J}_{2+}$ & $\hat{J}_{2-}$ 
& $\hat{\Theta}_n$ \\*[1mm] 
$TO$ & $M_3$ & ${\bf C}_3$ & $J_{3+}$ & $J_{3-}$ & $\Psi_n$ \\*[1.5mm]
\hline\hline
\end{tabular}
\end{center}

For convenience, in Table 1, we present the 
connection between the generic symbols, the operators, and states 
for each class of Schr\"odinger equation.  Recall that the eigenvectors
$\Phi_n$ and $\hat{\Theta}_n$ are related by 
$\hat{\Theta}(x,t)=R(\mu,\nu,\kappa)\Phi(x,t)$ 
[see Eqs. (I-36) and (I-65)] 
while $\hat{\Theta}_n$ and $\Psi_n$ are connected through 
$\Psi(x,t')=(\hat{\Theta}\circ t)(t')$ 
[see Eqs. (I-50) and (I-65)].

 
\section{Coherent States}

\subsection{The Displacement Operator}

In this and the following section, we shall continue to use generic 
symbols where convenient.  
In addition, we shall write the operators $J_{\pm}$ as 
\begin{equation}
J_- = i\left\{G_PP-G_XX+G_II\right\},~~~~~
J_+ = i\left\{-\bar{G}_PP+\bar{G}_XX-\bar{G}_II\right\},
\label{e:cs8}
\end{equation}
where the functions $G_P$, $G_X$, and $G_I$ are given in Table 2.  We 
have used Eq. (\ref{e:tqh8a}) for the
\begin{center}
\begin{tabular}{|c|c|c|c|}
\multicolumn{4}{l}{Table 2. Generic functions and their values according 
to class.}\\*[.2cm]
\hline\hline
\multicolumn{1}{|c|}{} & \multicolumn{1}{c|}{} & \multicolumn{1}{c|}{} & 
\multicolumn{1}{c|}{} \\*[-.2cm]
\multicolumn{1}{|c|}{Function} & \multicolumn{1}{c|}{$TO$}
& \multicolumn{1}{c|}{$TM$} & \multicolumn{1}{c|}{$TQ$}\\*[.2cm]\hline
  &  &  & \\*[-.2cm]
$G_P$ & $\xi(t)$  & $\hat{\xi}(t)$ & $\hat{\xi}(t)e^{\nu}+
2\hat{\dot{\xi}}(t)\kappa e^{-\nu}$ \\*[.2cm]\hline
  &  &  & \\*[-.2cm]
$G_X$ & $\dot{\xi}(t)$ & $\hat{\dot{\xi}}(t)$ & $\hat{\dot{\xi}}(t)e^{-\nu}$ 
\\*[.2cm]\hline
  &  &  & \\*[-.2cm]
$G_I$ & ${\cal C}(t)$  & $\hat{\cal C}(t)$    & $\hat{\cal C}+
\mu\hat{\dot{\xi}}(t)$\\*[.2cm]\hline
  &  &  & \\*[-.2cm]
$F_P$ & $\xi(t)\bar{\cal C}(t)-\bar{\xi}(t){\cal C}(t)$ & 
$\hat{\xi}(t)\hat{\bar{\cal C}}(t)-\hat{\bar{\xi}}(t)\hat{\cal C}(t)$ & 
$\left(\hat{\xi}(t)\hat{\bar{\cal C}}(t)-\hat{\bar{\xi}}(t)\hat{\cal C}(t)
\right)e^{\nu}-i\mu e^{\nu}$\\
  &  &  & $~~~~+2\left(\hat{\dot{\xi}}(t)\hat{\bar{\cal C}}(t)
-\hat{\dot{\bar{\xi}}}(t)\hat{\cal C}(t)\right)e^{-\nu}\kappa$\\*[.3cm]
\hline
  &  &  & \\*[-.2cm]
$F_X$ & $\dot{\xi}(t)\bar{\cal C}(t)-\dot{\bar{\xi}}(t){\cal C}(t)$ &
$\hat{\dot{\xi}}(t)\hat{\bar{\cal C}}(t)-\hat{\dot{\bar{\xi}}}(t)
\hat{\cal C}(t)$ & $\left(\hat{\dot{\xi}}(t)\hat{\bar{\cal C}}(t)
-\hat{\dot{\bar{\xi}}}(t)\hat{\cal C}(t)\right)e^{\nu}$ \\*[.3cm]  
\hline\hline 
\end{tabular}
\end{center}
definitions of the TQ functions.  
For convenience, we have dropped 
the prime on the variable $t$, since we 
do not make explicit use of the relationship between $t'$ and 
$t$ in this and the following section. 
The coefficients, $G_P$ and $G_X$, satisfy the relationship 
\begin{equation}
G_P\bar{G}_X-\bar{G}_PG_X = -i. \label{e:cs12}
\end{equation}
In essence, this contains expressions (\ref{e:to9}), (\ref{e:tme10}), 
and (\ref{e:tqh10}).

With this background, we define generic displacement-operator 
coherent states (DOCS), $\Omega_{\alpha}$, for $os(1)$-type systems 
in the usual way \cite{gt1}-\cite{amp1}: 
\begin{equation}
\Omega_{\alpha} = D(\alpha)\Omega_0. \label{e:cs16}
\end{equation}
$\alpha$ is a complex parameter, $\Omega_0$ from Table 1 is a 
generic extremal state, and  $D(\alpha)$ is a generic displacement 
operator  
\begin{equation}
D(\alpha) = \exp{\left[\alpha J_+-\bar{\alpha} J_-\right]},
\label{e:cs20}
\end{equation}
$D(\alpha)$ is unitary since $J_-$ 
and $J_+$ are Hermitian conjugates and $\alpha$ is a 
complex parameter.

By expressing the operators $x=X$ and $-i\partial_x=P$ in terms 
of $J_-$ and $J_+$, we can compute the expectation values in 
the usual way.  Using Eqs. (\ref{e:cs8}) to (\ref{e:cs12}), we 
find that 
\begin{eqnarray}
X & = & \bar{G}_PJ_-+G_PJ_++iF_PI,\label{e:cs24a}\\*[1mm]
P & = & \bar{G}_XJ_-+G_XJ_++iF_XI.\label{e:cs24b}
\end{eqnarray}
The purely imaginary functions $F_P$ and $F_X$ of Table 2 are defined as
\begin{equation}
F_P=G_P\bar{G}_I-\bar{G}_PG_I,~~~~~F_X=G_X\bar{G}_I
-\bar{G}_XG_I, \label{e:cs25}
\end{equation}
and specific values of these two functions for the three classes of systems 
are given in Table 2.


\subsection{Position and Momentum Expectation Values}

To calculate expectation values for position and momentum  we have  
\begin{eqnarray}
\langle x(t)\rangle & = &  \langle \Omega_{\alpha}|X| 
                            \Omega_{\alpha}\rangle  
    =  \langle \Omega_0|D^{-1}(\alpha) XD(\alpha)|\Omega_0\rangle 
    =  \alpha\bar{G}_P+\bar{\alpha}G_P+iF_P,   \label{e:cs40a}  \\
\langle p(t)\rangle & = & \langle \Omega_{\alpha}|P|
                           \Omega_{\alpha}\rangle 
     =  \langle \Omega_0|D^{-1}(\alpha)
                            PD(\alpha)|\Omega_0\rangle 
     =  \alpha\bar{G}_X+\bar{\alpha}G_X+iF_X.\label{e:cs40b}
\end{eqnarray}
To evaluate  $D^{-1}(\alpha)XD(\alpha)$ and $D^{-1}(\alpha)PD(\alpha)$
in Eqs. (\ref{e:cs40a}) and (\ref{e:cs40b}) we used the 
Eqs.  (\ref{e:cs24a}) and (\ref{e:cs24b}), the 
unitarity of $D(\alpha)$, Eq. (\ref{e:tqh1}), and the commutation 
relations (I-8) through (I-10). 

Let $x_o$ and $p_o$ be initial position and momentum:
\begin{equation}
x_o=\langle x(t_o)\rangle,~~~~~p_o =\langle p(t_o)\rangle.
\label{e:cs44}
\end{equation}
Placing a superscript `$o$' on $G_P$, $G_X$, $F_P$, and $F_X$ (and 
their corresponding values in Table 2) to indicate $t=t_o$, and 
using Eq. ({\ref{e:cs12}}), we find that
\begin{equation}
\alpha=i\left(G_P^op_o-G_X^ox_o\right)+G_P^oF_X^o-G_X^oF_P^o.
\label{e:cs52a}
\end{equation}
Substituting Eq. (\ref{e:cs52a}) for $\alpha$ into Eqs. (\ref{e:cs40a}) 
and (\ref{e:cs40b}), we obtain expressions for the expectation values 
of $X$ and $P$ in terms of $x_o$ and $p_o$:
\begin{eqnarray}
\langle x(t)\rangle & = & i\left(\bar{G}_PG_P^o-G_P
\bar{G}_P^o\right)p_o+i\left(G_P\bar{G}_X^o-\bar{G}_PG_X^o
\right)x_o\nonumber\\
 &   &  +iG_P\left(\bar{G}_I-\bar{G}_I^o\right)
-i\bar{G}_P\left(G_I-G_I^o\right),\label{e:cs56a}
\end{eqnarray}
\begin{eqnarray}
\langle p(t)\rangle & = & i\left(\bar{G}_XG_P^o-G_X
\bar{G}_P^o\right)p_o+i\left(G_X\bar{G}_X^o-\bar{G}_XG_X^o
\right)x_o\nonumber\\
 &   &  +iG_X\left(\bar{G}_I-\bar{G}_I^o\right)
-i\bar{G}_X\left(G_I-G_I^o\right),
\label{e:cs56b}
\end{eqnarray}
where we have used the definitions in Eqs. 
(\ref{e:cs12}) and (\ref{e:cs25}).

\indent For each of the three classes, we can combine Eqs. 
(\ref{e:cs56a}), and (\ref{e:cs56b}) with the functions in Table 2 to 
obtain explicit expectation values:  
\begin{eqnarray}
TQ\,:~~~\langle x(t)\rangle & = & i\left[\left(\hat{\bar{\xi}}\hat{\xi}\,^o
-\hat{\xi}\hat{\bar{\xi}}\,^o\right)e^{\nu}+2\left(\hat{\dot{\bar{\xi}}}
\hat{\xi}\,^o-\hat{\dot{\xi}}\hat{\bar{\xi}}\,^o\right)\kappa e^{-\nu}
\right]p_o\nonumber\\
 &   & +i\left[\left(\hat{\xi}\hat{\dot{\bar{\xi}}}\,^o-\hat{\bar{\xi}}
\hat{\dot{\xi}}\,^o\right)e^{\nu}+2\left(\hat{\dot{\xi}}
\hat{\dot{\bar{\xi}}}\,^o-\hat{\dot{\bar{\xi}}}\hat{\dot{\xi}}\,^o
\right)\kappa e^{-\nu}\right]x_o\nonumber\\
 &   & +\left[i\left(\hat{\xi}\hat{\bar{c}}-\hat{\bar{\xi}}\hat{c}
\right)+\mu\right]e^{\nu}+2i\left(\hat{\dot{\xi}}\hat{\bar{c}}
-\hat{\dot{\bar{\xi}}}\hat{c}\right)\kappa e^{-\nu},\nonumber\\*[1mm]
\langle p(t)\rangle & = & i\left(\hat{\dot{\bar{\xi}}}
\hat{\xi}\,^o-\hat{\dot{\xi}}\hat{\bar{\xi}}\,^o\right)e^{-\nu}p_o
+i\left(\hat{\dot{\xi}}\hat{\dot{\bar{\xi}}}\,^o-\hat{\dot{\bar{\xi}}}
\hat{\dot{\xi}}\,^o\right)e^{-\nu}x_o+i\left(\hat{\dot{\xi}}\hat{\bar{c}}
-\hat{\dot{\bar{\xi}}}\hat{c}\right)e^{-\nu}.\label{e:cs60c} \\
TM\,:~~~
\langle x(t)\rangle & = & i\left(\hat{\bar{\xi}}\hat{\xi}\,^o
-\hat{\xi}\hat{\bar{\xi}}\,^o\right)p_o
+i\left(\hat{\xi}\hat{\dot{\bar{\xi}}}\,^o-\hat{\bar{\xi}}
\hat{\dot{\xi}}\,^o\right)x_o+i\left(\hat{\xi}\hat{\bar{c}}
-\hat{\bar{\xi}}\hat{c}\right),\nonumber\\*[1mm]
\langle p(t)\rangle & = & i\left(\hat{\dot{\bar{\xi}}}
\hat{\xi}\,^o-\hat{\dot{\xi}}\hat{\bar{\xi}}\,^o\right)p_o
+i\left(\hat{\dot{\xi}}\hat{\dot{\bar{\xi}}}\,^o-
\hat{\dot{\bar{\xi}}}\hat{\dot{\xi}}\,^o\right)x_o
+i\left(\hat{\dot{\xi}}\hat{\bar{c}}-\hat{\dot{\bar{\xi}}}\hat{c}
\right).\label{e:cs60b} \\
TO\,:~~~
\langle x(t)\rangle & = & i\left(\bar{\xi}\xi^o-\xi\bar{\xi}^o
\right)p_o+i\left(\xi\dot{\bar{\xi}}\,^o-\bar{\xi}\dot{\xi}\,^o
\right)x_o+i\left(\xi\bar{c}-\bar{\xi}c\right),\nonumber\\*[1mm]
\langle p(t)\rangle & = & i\left(\dot{\bar{\xi}}\xi^o-\dot{\xi}
\bar{\xi}^o\right)p_o+i\left(\dot{\xi}\dot{\bar{\xi}}\,^o-
\dot{\bar{\xi}}\dot{\xi}\,^o\right)x_o+i\left(\dot{\xi}\bar{c}
-\dot{\bar{\xi}}c\right).\label{e:cs60a} 
\end{eqnarray}


\subsection{Uncertainties}

Now, we compute the uncertainty product for the general case. 
If we take the uncertainty of an operator A as 
\begin{equation}
(\Delta A)^2=\langle A^2\rangle-\langle A\rangle^2,\label{e:cs64}
\end{equation}
then we find that
\begin{equation}
(\Delta x)^2= G_P\bar{G}_P,~~~~~~(\Delta p)^2=G_X\bar{G}_X,
\label{e:cs68}
\end{equation}
which are both real and positive quantities.  Therefore, the 
uncertainty product has the form
\begin{equation}
(\Delta x)^2(\Delta p)^2  =  G_P\bar{G}_PG_X\bar{G}_X 
  =  \lfrac{1}{4}\left[1+\left(G_P\bar{G}_X+\bar{G}_PG_X\right)^2
\right].\label{e:cs72}
\end{equation}
This is both real and always greater than or equal to $1/4$.  (We 
used Eq. (\ref{e:cs12}) in the calculation of the second equality.)

We delay presentations of the particular 
uncertainties and uncertainty products for each of the 
classes of systems until the end of the corresponding section for 
squeezed states.  






\section{Squeezed States}

\subsection{The Squeeze Operator}

Define the operators {\cite{nt2}} 
\begin{equation}
{\cal K}_-=\lfrac{1}{2}J_-^2,~~~~~{\cal K}_+=\lfrac{1}{2}J_+^2,
~~~~~{\cal K}_3=J_+J_-+\lfrac{1}{2},\label{e:ss1}
\end{equation}
which satisfy the commutation relations {\cite{nt1a}}
\begin{equation}
[{\cal K}_+,{\cal K}_-]=-{\cal K}_3,~~~~~~[{\cal K}_3,
{\cal K}_{\pm}]=\pm 2{\cal K}_{\pm}.\label{e:ss4}
\end{equation}
Calculating their commutation relations with $J_{\pm}$, 
and using Eq. (\ref{e:eig1}), we find that 
\begin{eqnarray}
 & [{\cal K}_-,J_-]=0,~~~[{\cal K}_+,J_-]=-J_+,~~~~
[{\cal K}_3,J_-]=-J_-, & \nonumber\\*[1mm]
 & [{\cal K}_-,J_+]=J_-,~~~[{\cal K}_+,J_+]=0,~~~~
[{\cal K}_3,J_+]=+J_+. & \label{e:ss8} 
\end{eqnarray}
The algebra of operators, $\{{\cal K}_{\pm},{\cal K}_3,J_{\pm},I\}$, 

has the $su(1,1)\Box w_1^c$ structure.   

We define a generalized displacement-operator 
squeezed state {\cite{nt2,fns1,nt3,nt4}}, 
$\Omega_{\alpha,z}$, as follows: 
\begin{equation}
\Omega_{\alpha,z} = D(\alpha)S(z)\Omega_0,
\label{e:ss12}
\end{equation}
where $D(\alpha)$ is given in Eq. (\ref{e:cs20}) and $S(z)$ is the 
squeeze operator 
\begin{equation}
S(z)=\exp{(z{\cal K}_+-\bar{z}{\cal K}_-)}.\label{e:ss16}
\end{equation}
The parameter $z$ is complex.  For computational purposes, 
it is more convenient to write the squeeze operator in the form of 
``canonical coordinates of the second kind" \cite{nt2}.  A
Baker-Campbell-Hausdorff relation {\cite{nt2,fns1,drt4}} gives
this form as 
\begin{equation}
S(z)=\exp{(\gamma_+{\cal K}_+)}\exp{(\gamma_3{\cal K}_3)}
\exp{(\gamma_-{\cal K}_-)},\label{e:ss24}
\end{equation}
where 
\begin{equation}
\gamma_-=-{{\bar{z}}\over{|z|}}\tanh{|z|},~~~~~
\gamma_+= {{z}\over{|z|}}\tanh{|z|},~~~~~
\gamma_3=-\ln{\left(\cosh{|z|}\right)}, \label{e:ss28}
\end{equation}
\begin{equation}
z=re^{i\theta}, ~~~~~ r=|z|.   \label{e:ss30}
\end{equation}


\subsection{Position and Momentum Expectation Values}

To compute expectation values of position and momentum, we again 
employ the operators (\ref{e:cs24a}) and (\ref{e:cs24b}) and 
follow the same method of calculation as in Ref. {\cite{nt2}}.
\begin{equation}
\langle x(t)\rangle = \langle\alpha,z|X|\alpha,z\rangle =
\langle 0|S^{-1}(z)D^{-1}(\alpha)|X|D(\alpha)S(z)|0\rangle,
\label{e:ss32a}
\end{equation}
\begin{equation}
\langle p(t)\rangle = \langle\alpha,z|P|\alpha,z\rangle =
\langle 0|S^{-1}(z)D^{-1}(\alpha)|P|D(\alpha)S(z)|0\rangle.
\label{e:ss32b}
\end{equation}
Making use of Eq. (\ref{e:tqh1}) we obtain the adjoint action of 
the group operators $S(z)$ and $D(\alpha)$ on $X$ and $P$ 
respectively,
\begin{eqnarray}
S^{-1}(z)D^{-1}(\alpha)XD(\alpha)S(z) & = &
{\cal X}_{X,-}J_-+{\cal X}_{X,+}J_++{\cal X}_{X,I}I,\label{e:ss36a}
\\*[1mm]
S^{-1}(z)D^{-1}(\alpha)PD(\alpha)S(z) & = &
{\cal X}_{P,-}J_-+{\cal X}_{P,+}J_++{\cal X}_{P,I}I.\label{e:ss36b}
\end{eqnarray}
The coefficients of the operators in these two expressions are
\begin{eqnarray}
 & {\cal X}_{X,-}=\bar{G}_P\left(e^{\gamma_3}-\gamma_-\gamma_+
e^{-\gamma_3}\right)-G_Pe^{-\gamma_3} =\bar{G}_P\cosh{r}
+G_Pe^{-i\theta}\sinh{r}, & \nonumber\\*[1mm]
 & {\cal X}_{X,+}=\bar{G}_P\gamma_+e^{-\gamma_3}+G_Pe^{-\gamma_3}
= \bar{G}_Pe^{i\theta}\sinh{r}+G_P\cosh{r}, & \nonumber\\*[1mm]
 & {\cal X}_{X,I}=\alpha\bar{G}_P+\bar{\alpha}G_P+iF_P, & 
\label{e:ss40a}
\end{eqnarray}
\begin{eqnarray}
 & {\cal X}_{P,-}=\bar{G}_X\left(e^{\gamma_3}-\gamma_-\gamma_+
e^{-\gamma_3}\right)-G_Xe^{-\gamma_3} =\bar{G}_X\cosh{r}
+G_Xe^{-i\theta}\sinh{r}, & \nonumber\\*[1mm]
 & {\cal X}_{P,+}=\bar{G}_X\gamma_+e^{-\gamma_3}+G_Xe^{-\gamma_3}
= \bar{G}_Xe^{i\theta}\sinh{r}+G_X\cosh{r}, & \nonumber\\*[1mm]
 & {\cal X}_{P,I}=\alpha\bar{G}_X+\bar{\alpha}G_X+iF_X, & 
\label{e:ss40b}
\end{eqnarray}
where $G_X$ and $G_P$ are given in Table 2.

Combining Eqs.  (\ref{e:ss32a}) -  (\ref{e:ss40b}), 
we obtain the equations (\ref{e:cs40a}) and (\ref{e:cs40b}) for 
$\langle x(t)\rangle$ and $\langle p(t)\rangle$, 
respectively.  Identifying an initial position and momentum, as 
in Eq. (\ref{e:cs44}), we end up with Eqs.  (\ref{e:cs56a}) 
and (\ref{e:cs56b}) for $\langle x(t)\rangle$ and 
$\langle p(t)\rangle$ in terms of $x_o$ and $p_o$.  The 
expectation values for position and momentum for each of the 
three classes of equations are given in Eqs. (\ref{e:cs60c})  
through (\ref{e:cs60a}).  


\subsection{Uncertainties}

To obtain the squeezed-state uncertainty products, we proceed 
in the same way as we did for the coherent states, but with the 
operators (\ref{e:ss36a}) and (\ref{e:ss36b}).   
The uncertainties in position and momentum are 
\begin{eqnarray}
(\Delta x)^2 & = & \lfrac{1}{2}\left(\bar{G}_P^2e^{i\theta}+G_P^2
e^{-i\theta}\right)\sinh{2r}+G_P\bar{G}_P\cosh{2r},
\label{e:ss11a} \\
(\Delta p)^2 & = & \lfrac{1}{2}\left(\bar{G}_X^2e^{i\theta}+G_X^2
e^{-i\theta}\right)\sinh{2r}+G_X\bar{G}_X\cosh{2r}.
\label{e:ss11b}
\end{eqnarray}
These are both real and positive since  
$G_Pe^{-i\theta/2}+\bar{G}_Pe^{i\theta/2}$ and 
$G_Xe^{-i\theta/2}+\bar{G}_Xe^{i\theta/2}$ are both real. 
Particular 
expressions  $(\Delta x)^2$ and $(\Delta p)^2$, for the $TQ$, $TM$, 
and $TM$ systems, can be obtained by using the values 
of  $G_P$ and $G_X$ in Table 2.  

After some manipulation, we obtain 
the following expression for the uncertainty product:
\begin{eqnarray}
\left(\Delta x\right)^2\left(\Delta p\right)^2 & = &
\lfrac{1}{4}\left\{1+\left[\left(G_P\bar{G}_X+\bar{G}_PG_X
\right)\cosh{2r}\right.\right.\nonumber\\
 &   & \left.\left.+\left(G_PG_Xe^{-i\theta}+\bar{G}_P\bar{G}_X
e^{i\theta}\right)\sinh{2r}\right]^2\right\}
\label{e:ss44}
\end{eqnarray}
As for the coherent states, we have used the identity (\ref{e:cs12}) 
to aid in obtaining this result.

Note that coefficients of $\cosh{2r}$ in Eqs. 
(\ref{e:ss11a}) and (\ref{e:ss11b}) and of $\cosh^2{2r}$ 
in Eq. (\ref{e:ss44}), are identical 
to their respective coherent-state expressions (\ref{e:cs68}) and 
(\ref{e:cs72}).  
Also, as expected, the uncertainties and uncertainty 
products for the coherent states can be reclaimed by setting the 
squeezing parameters $r=\theta=0$ in the above. 

Finally, from Table 2, the uncertainty products for the 
three classes of equations are:
\begin{eqnarray}
TQ\,:& &\left(\Delta x\right)^2\left(\Delta p\right)^2  = 
 \lfrac{1}{4}\left\{1+\lfrac{1}{4}\left[\left(\hat{\dot{\phi}}_3
+8\hat{\dot{\xi}}\hat{\dot{\bar{\xi}}}\kappa e^{-2\nu}\right)
\cosh{2r}\right.\right.\nonumber\\
 &  & ~~~~~~~~\left.\left.+\left([\hat{\dot{\phi}}_1
+4\hat{\dot{\xi}}\,^2\kappa 
e^{-2\nu}]e^{-i\theta}+[\hat{\dot{\phi}}_2+4
\hat{\dot{\bar{\xi}}}\,^2\kappa 
e^{-2\nu}]e^{i\theta}\right)\sinh{2r}\right]^2\right\}.
\label{e:ss60}
\end{eqnarray}
\begin{eqnarray}
TM\,:& &\left(\Delta x\right)^2\left(\Delta p\right)^2  = 
\lfrac{1}{4}\left\{1+\lfrac{1}{4}\left[\hat{\dot{\phi}}_3
\cosh{2r}+\left(\hat{\dot{\phi}}_1e^{-i\theta}+\hat{\dot{\phi}}_2
e^{i\theta}\right)\sinh{2r}\right]^2\right\}.
\label{e:ss56}
\end{eqnarray}
The functions $\hat{\dot{\phi}}_j(t)$, $j=1,2,3,$ 
are in Eq. (\ref{e:tme10}).
\begin{eqnarray}
TO\,:& &\left(\Delta x\right)^2\left(\Delta p\right)^2  =  
\lfrac{1}{4}\left\{1+\lfrac{1}{4}\left[\dot{\phi}_3\cosh{2r}
+\left(\dot{\phi}_1e^{-i\theta}+\dot{\phi}_2e^{i\theta}\right)
\sinh{2r}\right]^2\right\} .
\label{e:ss48}
\end{eqnarray}
The functions $\phi_j(t')$, $j=1,2,3,$ are in Eq. 
(\ref{e:to20}).  

For the harmonic oscillator {\cite{nt2}}, 
$\dot{\phi}_3=0$, $\dot{\phi}_1=i\exp{[2i\omega(t-t_o)]}$,
$\dot{\phi}_2=-i\exp{[-2i\omega(t-t_o)]}$. (Recall we  ignore 
primes for time in this section.) The uncertainty 
product becomes 
\begin{equation}
\left(\Delta x\right)^2\left(\Delta p\right)^2 =
\lfrac{1}{4}\left[1+\lfrac{1}{4}\left(s^2-\frac{1}{s^2}\right)^2
\sin^2{[2\omega(t-t_o)-\theta]}\right],~~~~s=\exp{r}.\label{e:ss52}
\end{equation}
This expression is well known.  See, e.g.,  Eq. (86) in Ref. 
{\cite{nt5}}.
\vspace{2mm}


\section{Discussion} 

\subsection{Uncertainty Relations}

We have been considering 
Heisenberg-Weyl algebras with $J_-$, $J_+$, and $I$ 
satisfying the appropriate commutation relations.  Now define 
\begin{equation}
{\cal X} = \frac{J_-+J_+}{\sqrt{2}},
          ~~~~~~~~~~~~~{\cal P} = \frac{J_--J_+}{i\sqrt{2}}.
\label{xp}
\end{equation}
Then, the coherent states we have defined [see Eqs. (\ref{e:cs16}) and 
(\ref{e:cs20})],  
\begin{equation}
|\alpha\rangle = D(\alpha)|0\rangle, ~~~~~~~~~~~~~~
          D(\alpha) = \exp[\alpha J_+ - \bar{\alpha} J_-],    \label{disp}
\end{equation}
minimize the Heisenberg uncertainty relation
\begin{equation}
(\Delta {\cal X})^2 (\Delta {\cal P})^2 \ge 1/4.   \label{hur}
\end{equation}

In addition, in Eq. (\ref{e:ss1}), 
we have have defined the  $su(1,1)$ algebra with 
${\cal K}_-$, ${\cal K}_+$, and ${\cal K}_3$ satisfying the appropriate 
commutation relations among themselves and with the HW algebra.  

Then the squeezed states [see Eq. (\ref{e:ss12})], 
\begin{equation}
|\alpha, z\rangle = D(\alpha)S(z)|0\rangle, ~~~~~~~~~~~~~~
          S(z) = \exp[z {\cal K}_-+ - \bar{z} {\cal K}_+],  \label{ss}
\end{equation}
minimize the Schr\"odinger-Robertson uncertainty relation
\begin{equation}
(\Delta {\cal X})^2 (\Delta {\cal P})^2 
\ge \frac{1}{4} 
+ \frac{1}{4}~|~\langle ~ \{ {\cal X} - \langle {\cal X} \rangle,~
              {\cal P} - \langle {\cal P} \rangle \} ~ \rangle~|^2,
\label{sur}
\end{equation}
where $\{~,~\}$ is the anticommutator \cite{nagel}.  

Note that $x$ and $p$ are {\it not} ${\cal X}$ and ${\cal P}$, but 
linear combinations of them and $I$, with multiplicative coefficients.  
Therefore, although the uncertainty products of $x$ and $p$ are correct
and physically relevant, 
they do not necessarily satisfy the equalities in Eqs. 
(\ref{hur}) and (\ref{sur}).  In fact, they tend not to, except possibly 
for particular times (such as $t=t_o$).  They do, however, often 
tend to be close to minimum uncertainties.  

\subsection{The Classical Motion} 

For coherent and squeezed states, $\langle x(t) \rangle$ and 
$\langle p(t) \rangle$ should obey the classical 
Hamiltonian equations of motion:
\begin{equation}
\dot{x}=\frac{\partial H}{\partial p},~~~~~~~~\dot{p}=
-\frac{\partial H}{\partial x},\label{hameq}
\end{equation}
The `dot' indicates differentiation by $t'$ for $TO$ systems 
and differentiation by $t$ for $TM$ and $TQ$ systems, and $t$ and 
$t'$ are related through Eq. (I-67).

The classical Hamiltonians associated with each class of 
Schr\"odinger equations are
\begin{eqnarray}
TO\,:~~~H & = & \frac{p^2}{2}+g^{(2)}(t')x^2
+g^{(1)}(t')x+g^{(0)}(t'), \label{clhamto}\\*[2mm]
TM\,:~~~\hat{H} & = & e^{-2\nu}\frac{p^2}{2}+f^{(2)}(t)x^2
+f^{(1)}(t)x+f^{(0)}(t), \label{clhamtm}\\*[2mm]
TQ\,:~~~H & = & [1+k(t)]\frac{p^2}{2}-\frac{h(t)}{2}xp+
-\frac{g(t)}{2}p+h^{(2)}(t)x^2+h^{(1)}(t)x+h^{(0)}(t). \label{clhamtq}
\end{eqnarray}
Putting these Hamiltonians into Eqs. (\ref{hameq}) one finds
\begin{eqnarray}
&TO\,:&~~~~~~\dot{x} = p,~~~~~~~~~
\dot{p} = -2g^{(2)}(t')x-g^{(1)}(t'),
\label{clheqto}\\*[2mm]
&TM\,:&~~~~~~\dot{x} = e^{-2\nu}x,~~~~~~~~~
\dot{p} = -2f^{(2)}(t)x-f^{(1)}(t), 
\label{clheqtm}\\*[2mm]
&TQ\,:&~~~~~~\dot{x} = [1+k(t)]p-\frac{h(t)}{2}x
  -\frac{g(t)}{2},\nonumber\\*[1mm]
&     &~~~~~~\dot{p}=-\frac{h(t)}{2}p
 - 2h^{(2)}(t)x -h^{(1)}(t). \label{clheqtq}
\end{eqnarray}

Now consider the expectation values $\langle x \rangle$ and 
$\langle p \rangle$ in  Eqs. (\ref{e:cs60c}) 
through (\ref{e:cs60a}) and their time derivatives.   
Making  extensive use  of Eqs. (I-32), (I-33), (I-59)-(I-61), (I-66), 
(\ref{e:to9}), (\ref{e:tme10}), and (\ref{e:tme12}), one can demonstrate 
that these quantities satisfy Eqs. (\ref{clheqto}) to (\ref{clheqtq}) with 
$x \rightarrow \langle x \rangle$ and 
$p \rightarrow \langle p \rangle$.  Thus, the classical motion is 
satisfied.  


\section*{Acknowledgements}

MMN acknowledges the support of the United States Department of 
Energy.  DRT acknowledges
a grant from the Natural Sciences and Engineering Research Council 
of Canada.



\begin{thebibliography}{99}

\bibitem{nt1} M. M. Nieto and D. R. Truax (previous article).  
Eprint quant-ph/981075..

\bibitem{drt1} D. R. Truax, 
J. Math. Phys. {\bf 22}, 1959-1964 (1981).

\bibitem{drt2} D. R. Truax, 
J. Math. Phys. {\bf 23}, 43-54 (1982).

\bibitem{kt1} A. Kalivoda and D. R. Truax, 
To be published.

\bibitem{nt2} M. M. Nieto and D. R. Truax, 
J. Math. Phys. {\bf 38}, 84-97 (1997).

\bibitem{gt1} S. Gee and D. R. Truax, 
Phys. Rev. A {\bf 29}, 1627-1638 (1984).

\bibitem{rjg1} R. J. Glauber, 
Phys. Rev. {\bf 131}, 2766-2788 (1963).

\bibitem{amp1} A. M. Perelomov, 
Commun. Math. Phys. {\bf 26}, 222-236 (1972).   

\bibitem{fns1} R. A. Fisher, M. M. Nieto, and V. D. Sandberg, 
Phys. Rev. D {\bf 29}, 1107-1110 (1984).

\bibitem{nt3} M. M. Nieto and D. R. Truax, 
Phys. Rev. Lett. {\bf 71}, 2843-2846 (1994).

\bibitem{else} M. M. Nieto and D. R. Truax (in preparation).

\bibitem{wm1} W. Miller, Jr., {\it Symmetry and Separation of 
Variables} (Addison-Wesley, Reading, MA, 1977).

\bibitem{wm2} W. Miller, Jr., {\it Symmetry Groups and 
their Applications} (Academic, New York, 1972).

\bibitem{nt1a}  In Ref. {\cite{nt2}}, there is a typographical 
error in Eq. (52).  The second commutator should read 
$[{\cal K}_3,{\cal K}_{\pm}]=\pm 2{\cal K}_{\pm}$.  See 
the present Eq. (\ref{e:ss4}).

\bibitem{nt4} M. M. Nieto and D. R. Truax, 
Fortschritte Phys. {\bf 45}, 145-156 (1997). 

\bibitem{drt4} D. R. Truax, 
Phys. Rev. D {\bf 31}, 1988-1991 (1985).

\bibitem{nt5} M. M. Nieto and D. R. Truax, 
J. Math. Phys. {\bf 38}, 98-114 (1997).

\bibitem{nagel} B. Nagel, {\it Higher power squeezed states, Jacobi 
Matrices, and the Hamburger moment problem.}  In: {\it Proceedings of the 
Fifth International Conference on Squeezed States and Uncertainty 
Relations}, NASA Conference Publication NASA/CP-1998-206855,  
ed. D. Han, J. Janszky, Y. S. Kim, and V. I. Man'ko  
(NASA, Washington, DC, 1998), pp. 43-48, and references therein.
Eprint quant-ph/9711028.


\end{thebibliography}
\end{document}